\documentclass[10pt]{article}
\begin{document}
\centerline {{\Large Approaches to general field theory}}  
\centerline  {{\large (The method of skew-symmetric differential forms)}}
\centerline {L.I. Petrova}
\centerline{{\it Moscow State University, Russia, e-mail: ptr@cs.msu.su}}
\bigskip

The basis for the field theory are properties of the closed exterior 
differential forms (skew-symmetric differential forms defined on manifolds with 
the closed metric forms), which reflect properties of the conservation laws for 
physical fields. It is possible to classify physical fields and interactions. 
So, the (0-form) corresponds to the strong interaction, the (1-form) 
corresponds to the weak interaction, the (2-form) coorresponds to the 
electromagnetic interaction, and the (3-form) corresponds to the gravitational 
interaction. This is the basis of unified field theory. 
 
As a general field theory it can be a theory, which not only decribes possible 
physical fields and relation between them, but also discloses a mechanism of 
forming physical fields and the causality of such processes. It occurs that as 
the basis of such a theory it can become the theory of skew-symmetric 
differential forms defined on manifolds with unclosed metric forms. These 
differential forms, which were named the evolutionary ones, reflrect the 
properties of the conservation laws for material media (the balance 
conservation laws for energy, linear and angular momentum, and mass) and 
disclose a mechanism of the evolutionary processes in material media. It is in 
such processes the physical structures that form physical fields originate. 
The theory of exterior and evolutionary skew-symmetric differential forms 
discloses the causality of physical processes, establishes a relation between 
physical fields and material media and allows to introduce a classification 
of physical fields and interactions. 

\section{The role of exterior differential forms in invariant field theories} 

The analysis of operators and equations of existing invariant field theories 
shows that the mathematical principles of the theory of closed exterior 
differential forms lie at the basis of existing field theories. 

A connection of field theory with the exterior differential forms is explained 
by the fact that the closed exterior differential forms describe physical 
structures, which constitute physical fields. This is connected with the 
conservation laws [1-3]. 

\subsection*{Properties of closed exterior differential forms, which reflect 
properties of the conservation laws }

The exterior differential form of degree $p$ ($p$-form) can be written as [4-6] 
$$
\theta^p=\sum_{i_1\dots i_p}a_{i_1\dots i_p}dx^{i_1}\wedge
dx^{i_2}\wedge\dots \wedge dx^{i_p}\quad 0\leq p\leq n\eqno(1.1)
$$
Here $a_{i_1\dots i_p}$ are the functions of the variables $x^{i_1}$,
$x^{i_2}$, \dots, $x^{i_p}$,  $n$ is the dimension of space,
$\wedge$ is the operator of exterior multiplication, $dx^i$,
$dx^{i}\wedge dx^{j}$, $dx^{i}\wedge dx^{j}\wedge dx^{k}$, \dots\
is the local basis which satisfies the condition of exterior
multiplication:
$$
\begin{array}{l}
dx^{i}\wedge dx^{i}=0\\
dx^{i}\wedge dx^{j}=-dx^{j}\wedge dx^{i}\quad i\ne j
\end{array}\eqno(1.2)
$$
[From here on the symbol $\sum$ will be omitted and it will be
implied that the summation is performed over double subscripts.  Besides, the
symbol of exterior multiplication will be also omitted for the
sake of presentation convenience].

The differential of the (exterior) form $\theta^p$ is expressed as
$$
d\theta^p=\sum_{i_1\dots i_p}da_{i_1\dots
i_p}dx^{i_1}dx^{i_2}\dots dx^{i_p} \eqno(1.3)
$$

From a definition of differential one can see that, firstly, the
differential of the exterior form is also the exterior form
(but with the degree $(p+1)$), and, secondly, he can see
that the components of the differential form commutator are the 
coefficients of the form differential. Thus, the differential
of the first-degree form $\omega=a_i dx^i$ can be written as
$d\omega=K_{ij}dx^i dx^j$ where $K_{ij}$ are the components of the
commutator for the form $\omega$ that are defined as
$K_{ij}=(\partial a_j/\partial x^i-\partial a_i/\partial x^j)$.

The theory of exterior differential forms was developed just for differentiable 
manifolds and manifolds with structures of any types. They may be the Hausdorff 
manifolds, fiber spaces, the comological, characteristical, configuration manifolds, 
and so on. These manifolds and their properties are treated in [6-8] and in 
some other works. Since all these manifolds possess
structures of any types, they have one common property, namely,
locally they admit one-to-one mapping into the Euclidean subspaces
and into other manifolds or submanifolds of the same dimension [8]. 
What they have in common is that the metric forms of such manifolds are closed. 
Below we will consider differential forms, which are defined on manifolds with 
metric forms that are unclosed. Differential of such forms will include an 
additional term that contains a differential of unclosed metric form. Such 
skew-symmetric differential forms, which were named the evolutionary ones, 
possess new unique possibilities that disclose properties of field theories. 

If $\theta^p$ be the exterior differential form of degree $p$ ($p$-form), 
the closure conditions of the exterior differential form (vanishing the form 
differential) can be written as
$$
d\theta^p=0\eqno(1.4)
$$
From this equation one can see that the closed form is a conservative 
quantity. This means that it can correspond to the conservation law, namely, 
to some conservative physical quantity. 

In relation (1.4) the exterior differential form is an exact one. 
If the exterior differential form is closed only on pseudostructure, 
that is, this form is a closed {\it inexact} differential form, the closure
condition is written as
$$
d_\pi\theta^p=0\eqno(1.5)
$$
And the pseudostructure $\pi$ obeys the condition
$$
d_\pi{}^*\theta^p=0\eqno(1.6)
$$
where ${}^*\theta^p$ is the dual form. 

From conditions (1.5) and (1.6) one can see that the exterior differential 
form closed on pseudostructure is a conservative object, namely, this
quantity conserves on pseudostructure. This can also correspond to
some conservation law, i.e. to conservative object.

The closure conditions for the exterior differential
form ($d_{\pi }\,\theta ^p\,=\,0$)
and the dual form ($d_{\pi }\,^*\theta ^p\,=\,0$) are
mathematical expressions of the conservation law. 
Such conservation laws that state the existence of
conservative physical quantities or objects can be named the exact ones.   

The pseudostructure and the closed exterior form defined on 
the pseudostructure make up a binary differential and geometrical structure. 
Such a binary object can be named a Bi-Structure. 
(This is an example of the differential and geometrical structure (G-Structure).) 
It is evident that such a structure does correspond to the conservation law.

The physical structures, from which physical fields are formed, are precisely 
structures that correspond to the exact conservation law.

Relations that define the physical structures ($d_{\pi }\,\theta ^p\,=\,0$,
$d_{\pi }\,^*\theta ^p\,=\,0$) turn out to be coincident with the mathematical
expression for the exact conservation law.

The mathematical expression for the exact conservation law and its connection
with physical fields can be schematically written in the following way
$$
\def\\{\vphantom{d_\pi}}
\cases{d_\pi \theta^p=0\cr d_\pi {}^{*\mskip-2mu}\theta^p=0\cr}\quad
\mapsto\quad
\cases{\\\theta^p\cr \\{}^{*\mskip-2mu}\theta^p\cr}\quad\hbox{---}\quad
\hbox{physical structures}\quad\mapsto\quad\hbox{physical fields}
$$

It is obvious that the exact conservation law is that for physical fields.

\subsection*{Characteristic properties of the closed exterior forms and their 
relation to properties of existing field theories} 

Since the relations for exact conservation laws and corresponding physical 
structures (which form physical fields) are expressed in terms of closed 
and dual forms, it is obvious that at the basis of all existing field theories 
(which describe physical fields) there lie properties of the closed exterior 
differential and dual forms. The properties and the mathematical apparatus of 
exterior differential forms allow to disclose  specific features peculiar to 
all existing field theories.

1) {\it Invariance of closed exterior forms} 

From the closure condition of exterior differential it follows a 
property of exterior differential forms, which has a physical meaning, namely, 
any  closed exterior form is a differential of the form 
of lower degree: the total one if the form is exact
$$\theta^p=d\theta^{p-1}\eqno(1.7)$$ 
or the interior one on pseudostructure if the form is inexact
$$\theta^p=d_\pi\theta^{p-1}\eqno(1.8)$$  

Since the closed exterior form is a differential then it is
obvious that the closed form proves to be invariant under all
transformations that conserve the differential. The unitary 
transformations (0-form), the tangent and canonical transformations (1-form), 
the gradient and gauge transformations (2-form) and so on are  examples of such 
transformations. {\it These are gauge transformations for spinor, 
scalar, vector, tensor (3-form) fields}.  

It is well known that these are transformations typical for existing 
field theories. The equations of existing field theories remain invariant under 
such transformations.

At this point it should be emphasized that the relation between the closed 
exterior form and the form of lower degree shows that the form of lower 
degree can correspond to the 
potential, and the closed form by itself can correspond to the potential force. 

2) {\it Conjugacy of the closed exterior forms}

The closure of the exterior differential forms and hence their
invariance result from the conjugacy of elements of the exterior or dual forms.
On the other hand, the concept of conjugacy may imply something that leads to
the closure of exterior or dual forms, obeys the closure condition,
or establishes a relation between closed forms.

From the definition of the exterior differential form
one can see that the exterior differential forms have complex structure. 
The specific features of the exterior form structure are a homogeneity with
respect to the basis, skew-symmetry, the integration of terms
each consisting of two objects of different nature
(the algebraic nature for the form coefficients, and the geometric nature
for the base components). Besides,  the exterior form depends
on the space dimension and on the manifold topology. The closure
property of the exterior form means that any objects, namely,
elements of the exterior form, components of elements, elements of
the form differential, exterior and dual forms, and others, turn
out to be conjugated. The variety of objects of conjugacy leads to 
the fact that the closed forms can describe a great number of different 
physical and spatial structures.
It is the conjugacy that leads to realization of
the invariant and covariant properties of the exterior and dual
forms. These properties of exterior differential forms lie just at the basis 
of field theories.

2) {\it Identical relations of the closed exterior forms}

Since the conjugacy is a certain connection between two operators or
mathematical objects, it is evident that relations can be used to express
conjugacy mathematically. Just such relations, which are the identical one, 
constitute the basis of the mathematical apparatus of the exterior 
differential forms.

The identical relations for exterior differential forms reflect the
closure conditions of the differential forms, namely, vanishing the form
differential (see formulas (1.4), (1.5), (1.6)) and the conditions
connecting the forms of consequent degrees (see formulas (1.7), (1.8)).
Hence they are a mathematical expression of the conservation laws (which correspond 
to physical structures forming physical fields) and a mathematical expression 
of the invariance and covariance. 
And this lies at the basis of existing field theories.

One can assure himself that all existing field theories contain the identical 
relations, which are the identical relations of the exterior differential forms, 
or their differential or integral representations.

Examples of such relations are canonical 
relations in the Schr\H{o}dinger equations, gauge invariance in electromagnetic 
theory, commutator relations in the Heisenberg theory, 
symmetric connectednesses,  
identity relations by Bianchi in the Einstein theory, cotangent bundles in 
the Yang-Mills theory, the covariance conditions in the tensor 
methods, the characteristic relations (integrability conditions) in equations 
of mathematical physics, etc. 

\subsection*{Characteristical properties and peculiarities of existing field 
theories} 

A connection between the exterior differential forms and existing field theories 
allow to disclose peculiarities of the field theory equations, their common 
functional properties and their interconnection.

Practically all field theory operators are expressed 
in terms of following operators of the exterior differential forms: 
$d$ (exterior differential), $\delta$ (the operator of transforming the form 
of degree $p+1$ into the form of degree $p$), $\delta '$ (the operator of 
cotangent transformations), $\Delta $ (that of the transformation $d\delta-\delta d$), 
$\Delta '$ (the operator of the transformation $d\delta'-\delta' d$). 
In terms of these operators that act onto exterior forms one can write down the 
operators by Green, d'Alembert, Laplace and the operator of canonical 
transformations [9,10]. Eigenvalues of these operators reveal themselves as conjugacy 
conditions for the differential form elements. 

The equations, that are equations of the existing field theories, are those
obtained on the basis of the properties of the exterior differential form 
theory. To the equations of quantum mechanics (equations by Shr\H{o}dinger, 
Heisenberg, Dirac) there correspond the closed exterior forms of zero degree 
or appropriate dual forms. The closed exterior form of zero degree corresponds 
to the Schr\H{o}dinger 
equation, the close dual form corresponds to the Heisenberg equation. 
It can be pointed out that, whereas the equations by Shr\H{o}dinger 
and Heisenberg describe a behavior of potential obtained from the zero 
degree closed form, Dirac's {\it brac-} and {\it ket}- vectors 
constitute the zero degree closed exterior form itself as the result of 
conjugacy (vanishing the scalar product).

The Hamilton formalism is based on the properties of closed exterior and dual 
forms of the first degree. The closed exterior differential form 
$ds=-Hdt+p_j dq_j$ (the Poincare invariant) corresponds to field equation [10]. 

The properties of closed exterior and dual forms of the second 
degree lie at the basis of the electromagnetic field equations. The Maxwell 
equations may be written as  
$d\theta^2=0$, $d^*\theta^2=0$ [9], where $\theta^2=
\frac{1}{2}F_{\mu\nu}dx^\mu dx^\nu$ (here $F_{\mu\nu}$ is the strength tensor). 

Closed exterior and dual forms of the third degree correspond to the 
gravitational field. 

The connection between field theory and  closed exterior differential forms 
supports the invariance of field theory.

The invariance of field theories is an invariance under transformations that 
conserve the differential. These are transformations under which the invariance 
of closed exterior forms is conserved. As it was already pointed out, these are 
the unitary transformations (0-form), the tangent and canonical transformations 
(1-form), the gradient and gauge transformations (2-form) are gauge 
transformations for tensor fields (3-form).  

The covariance of the dual form is directly connected with the invariance
of the exterior closed inexact form. The covariance of the dual form
play an important role in describing physical structures and manifolds.

And here it should underline that the field theories are based on the properties 
of closed {\it inexact} forms. This is explained by the fact that only inexact 
exterior forms  can correspond to the physical structures that form 
physical fields.  The condition that the closed exterior forms, 
which constitute the basis of field theory equations, are inexact ones 
reveals in the fact that essentially all existing field theories include 
a certain elements of noninvariance, i.e. they are based either on functionals 
that are not identical invariants (such as Lagrangian, action functional, entropy) 
or on equations (differential, integral, tensor, spinor, matrix and so on) 
that have no identical invariance (integrability or covariance). Such elements 
of noninvariance are, for example, nonzero value of the curvature tensor in 
Einstein's theory [11], the indeterminacy principle in Heisenberg's theory, 
the torsion in the theory by Weyl [11], the Lorentz force in electromagnetic 
theory [12], an absence of general integrability of the Schr\H{o}dinger 
equations, 
the Lagrange function in the variational methods, an absence of the identical 
integrability of the mathematical physics equations and an absence of identical 
covariance of the tensor equations, 
and so on. Only if we assume elements of noncovariance, we can obtain 
closed {\it inexact} forms that correspond to physical structures.

And yet,  the existing field theories are invariant ones because they are 
provided with additional conditions under which the invariance or covariance 
requirements have to be satisfied. These 
conditions are the closure conditions of exterior or dual forms. 
Examples of such conditions are the above pointed identity relations: 
canonical, gauge, commutator relations, 
symmetric connectednesses, identity relations by Bianchi etc. 

From the aforesaid one can see that both the field theory transformations 
and the field theory equations (identical relations) as well are characterized 
by a degree of the closed form. This discloses a relation between them and 
shows that it is possible to introduce a classification of physical fields 
according to the degree of exterior differential form. 

As it will be shown below, such a classification is true also for physical 
interactions. If to denote the degree of closed exterior form by $k$, the case 
$k=0$ will correspond to the strong interaction, $k=1$ will do to the weak 
interaction,  $k=2$ will correspond to the electromagnetic interaction, and 
$k=3$ will correspond to the gravitational interaction.

But within the framework of only 
exterior differential forms one cannot understand how this classification 
is explained. This can be elucidated only by application of skew-symmetric 
differential forms of another type, which possess not invariant properties 
but evolutionary ones. Such differential forms are just skew-symmetric 
differential forms, which are defined on 
manifolds with nonclosed metric forms and were named  the 
evolutionary differential forms. 

\section{A role of evolutionary differential forms in  
field theory} 

In paper [2] it has been noted that one must distinguish two types 
of differential equations of mathematical physics: 

1) differential equations that describe physical processes, and 

2) the invariant equations of the field theory that describe 
physical structures forming physical fields.

As it has been shown above the field theories are based on the conservation 
laws. At the basis of the field theory equations there lie  properties of the 
skew-symmetric differential forms.

It turns out that differential equations, which describe physical processes,  
are also based on the conservation laws. And at the basis of these equations 
there also lie properties of the skew-symmetric differential forms. 

A difference between two types of equations of mathematical physics consists 
in the following. 

The conservation laws, on which field theories are based, are those 
{\it for physical fields}. The skew-symmetric differential forms correspond to 
{\it the closed exterior 
differential forms} (skew-symmetric differential forms defined on manifolds 
with closed metric forms).

In contrast to this, the conservation laws, on which differential equations 
that describe physical processes are 
based, are the conservation laws {\it for material media (material systems)}. 
And skew-symmetric 
differential forms correspond to {\it evolutionary differential forms} 
(skew-symmetric differential forms defined on manifolds with unclosed 
metric forms).

The connection between the mathematical physics equations and the skew-symmetric 
differential forms enables one to see a connection of the field theory equations 
with equations that describe physical processes. And this, in turn, allows 
to see an internal connection of existing invariant field theories and 
a validity of these theories.

Here it should be noted some functional peculiarities of differential 
equations.

In differential equations of mathematical physics, which describe physical 
processes, the functions are found by means of integration of derivatives 
obtained from the differential equation. Whereas in the field theory equations 
the functions are obtained not from derivatives, but from differentials, and 
they are  exterior forms (potentials or the state functions). And differentials 
themselves are closed forms, i.e. they are invariants.

Differential equations of mathematical physics, which describe physical 
processes in material media, in addition to operators (derivatives) of the 
functions desired, involve the terms that are connected with an external 
action on the system under consideration. Such terms cannot be invariant ones. 
Hence, these differential equations cannot be invariant equations. 

A peculiarity of the invariant equations consists in that they involve only 
functions or operators on the functions desired. Due to this fact they can 
be reducted to identical relations or are identical relations.

To such identical relations it can be reduced the field theory equations, 
for example, the Maxwell equations, Einstein's equations, the Schr\H{o}dinger 
equation, Dirac's equation, and so on. 

So, the Maxwell equations are reduced to the forms $\theta^2=0 $ 
and $*\theta^2=0 $, where  $\theta^2 $ is the second degree form. Field 
equation [2] is reduced to the canonical relations that corresponds 
to the closure condition of the dual form and the first degree exterior form. 
The Schr\H{o}dinger equation is an analog of 
the field equation for zero degree form. Einstein's equation is an identical 
relation. This equation connects a differential of the first degree form 
and the closed form of the second degree, namely, the energy-momentum tensor. 
(It would we noted that, though Einstein's equation connects the closed forms 
of the second degree, this equation follows from the third degree differential 
forms [13]).   

Thus, we obtain that differential equations are connected with relations.
  
It appears that noninvariant differential equations are also connected with 
relations. However, in contrast to invariant equations, which are connected 
with identical relations, noninvariant differential equations are connected 
with {\it nonidentical} relations.

Relations, with which differential equations are connected, are expressed in 
terms skew-symmetric differential forms. In this case identical relations are 
expressed in terms of closed exterior forms (as it has been shown above), and 
the nonidentical relations involve the unclosed form.

As it will be shown below, differential equations that describe physical 
processes are convolved into nonidentical relations. From such nonidentical 
relations it can be obtained  identical relations of the closed exterior forms 
that lie at the basis of invariant equations of field theory.

Nonidentical relations are those that involve skew-symmetric 
differential forms defined on manifolds with metric forms - evolutionary 
differential forms.   

\subsection*{Some properties of evolutionary  differential forms}

As it was already mentioned, the evolutionary differential forms are skew-sym\-metric 
differential forms defined on manifolds with metric forms that are unclosed. 
The evolutionary differential form of degree $p$ ($p$-form),
as well as the exterior differential form, can be written down as
$$
\omega^p=\sum_{\alpha_1\dots\alpha_p}a_{\alpha_1\dots\alpha_p}dx^{\alpha_1}\wedge
dx^{\alpha_2}\wedge\dots \wedge dx^{\alpha_p}\quad 0\leq p\leq n\eqno(2.1)
$$
where the local basis obeys the condition of exterior multiplication
$$
\begin{array}{l}
dx^{\alpha}\wedge dx^{\alpha}=0\\
dx^{\alpha}\wedge dx^{\beta}=-dx^{\beta}\wedge dx^{\alpha}\quad
\alpha\ne \beta
\end{array}
$$
(summation over repeated subscripts is implied).

But the evolutionary form differential cannot be written similarly to that 
presented for exterior differential forms (see formula (1.3)). In the 
evolutionary form differential there appears an additional term connected with 
the fact that the basis of the form changes. For the differential forms defined 
on the manifold with unclosed metric form one has 
$d(dx^{\alpha_1}dx^{\alpha_2}\dots dx^{\alpha_p})\neq 0$ 
(it should be noted that for differentiable manifold the following is valid:   
$d(dx^{\alpha_1}dx^{\alpha_2}\dots dx^{\alpha_p}) = 0$).  
For this reason a differential of the evolutionary form $\omega^p$ can be 
written as 
$$
d\omega^p{=}\!\sum_{\alpha_1\dots\alpha_p}\!da_{\alpha_1\dots\alpha_p}dx^{\alpha_1}dx^{\alpha_2}\dots
dx^{\alpha_p}{+}\!\sum_{\alpha_1\dots\alpha_p}\!a_{\alpha_1\dots\alpha_p}d(dx^{\alpha_1}dx^{\alpha_2}\dots
dx^{\alpha_p})\eqno(2.2)
$$
where the second term is connected with a differential of the basis. That
is expressed in terms of the metric form commutator[2]. For manifold with
closed metric form this term vanishes.

Every evolutionary form is unclosed form, since its commutator, and, 
conseguently, a differential of this form are nonzero (the evolutionary form 
commutator involves a commutator of unclosed metric form, which is nonzero).

In more detail about properties of the evolutionary forms and peculiarities 
of their mathematical apparatus it was written in work [2]. Here we shall call 
attention only to properties of the evolutionary forms that correspond to 
the conservation laws.

The evolutionary  differential forms, as well as the exterior  differential 
forms, can reflect properties of the conservation laws. However, in contrast to 
exterior differential forms, which reflect properties of the conservation laws 
for physical fields, the evolutionary differential forms reflect properties of 
the conservation laws for material systems (material media).

\{Material 
system is a variety of elements that have internal structure and interact 
to one another. As examples of material systems it may be thermodynamic, 
gas dynamical, cosmic systems, systems of elementary particles  
and others. Examples of elements that constitute a material system 
are electrons, protons, neutrons, atoms, fluid particles, cosmic objects, and 
others\}.

The conservation laws for material systems are the conservation laws for energy, 
linear momentum, angular momentum, and mass. These are conservation laws 
that can be named as balance conservation laws. In contrast to the conservation 
laws for physical fields, which state an existence of conservative 
physical quantities or objects, the conservation laws for material systems 
establish a balance between
a variation of physical quantity and the corresponding external action. 

In works [2,3] it has been shown that the balance conservation laws play a 
controlling role in the evolutionary processes, which lead to origination of 
physical structures. Mathematical apparatus of the evolutionary differential 
forms, that describe properties of the balance conservation laws is significant 
for understanding foundations of the general field theory.

\subsection*{Properties of evolutionary differential forms, which reflect 
properties of the balance conservation laws}

Let us analyze the equations
that describe the balance conservation laws for energy and linear momentum.

We introduce two frames of reference: the first is an inertial one
(this frame of reference is not connected with material system), and
the second is an accompanying
one (this system is connected with manifold constructed of 
trajectories of material system elements). The energy equation
in the inertial frame of reference can be reduced to the form:
$$
\frac{D\psi}{Dt}=A_1 \eqno(2.3)
$$
where $D/Dt$ is the total derivative with respect to time (or another 
evolutionary variable), $\psi $ is the functional
of the state that specifies a material system, $A_1$ is the quantity that
depends on specific features of the system and on external energy actions onto
the system. \{The action functional, entropy, wave function
can be regarded as examples of the functional $\psi $. Thus, the equation
for energy presented in terms of the action functional $S$ has a similar form:
$DS/Dt\,=\,L$, where $\psi \,=\,S$, $A_1\,=\,L$ is the Lagrange function.
In mechanics of continuous media the equation for
energy of ideal gas can be presented in the form [14]: $Ds/Dt\,=\,0$, where
$s$ is entropy. In this case $\psi \,=\,s$, $A_1\,=\,0$. It is worth noting 
that the examples presented show that the action functional and entropy 
play the same role.\}

In the accompanying frame of reference the total derivative with respect to
time is transformed into the derivative along trajectory. Equation (2.3)
is now written in the form
$$
{{\partial \psi }\over {\partial \xi ^1}}\,=\,A_1 \eqno(2.4)
$$
here $\xi^1$ is the coordinate along trajectory.

In a similar manner, in the
accompanying frame of reference the equation for linear momentum appears
to be reduced to the equation of the form 
$$
{{\partial \psi}\over {\partial \xi^{\nu }}}\,=\,A_{\nu },\quad \nu \,=\,2,\,...\eqno(2.5)
$$
where $\xi ^{\nu }$ are the coordinates in the direction normal to trajectory, 
$A_{\nu }$ are the quantities that depend on the specific
features of the system and external force actions.

Eqs. (2.4) and (2.5) can be convolved into the relation
$$
d\psi\,=\,A_{\mu }\,d\xi ^{\mu },\quad (\mu\,=\,1,\,\nu )\eqno(2.6)
$$
where $d\psi $ is the differential
expression $d\psi\,=\,(\partial \psi /\partial \xi ^{\mu })d\xi ^{\mu }$.

Relation (2.6) can be written as
$$
d\psi \,=\,\omega \eqno(2.7)
$$
Here $\omega \,=\,A_{\mu }\,d\xi ^{\mu }$ is the differential form of the
first degree.

Relation (2.7) was obtained from the equation of the balance
conservation laws for
energy and linear momentum. In this relation the form $\omega $ is that of the
first degree. If the equations of the balance conservation laws for
angular momentum be added to the equations for energy and linear momentum,
this form in the evolutionary relation will be the form of the second degree.
And in  combination with the equation of the balance conservation law
of mass this form will be the form of degree 3.

Thus, in the general case the evolutionary relation can be written as
$$
d\psi \,=\,\omega^p \eqno(2.8)
$$
where the form degree  $p$ takes the values $p\,=\,0,1,2,3$..
(The evolutionary
relation for $p\,=\,0$ is similar to that in the differential forms, and it 
was obtained from the interaction of energy and time.)

Since the equation of the balance conservation laws are the evolutionary ones, the relation
obtained is also an evolutionary relation.

The evolutionary relation is a nonidentical one as it involves the unclosed 
differential form. 

Let us consider the commutator of the
form $\omega \,=\,A_{\mu }d\xi ^{\mu }$.
Components of the commutator of such a form can
be written as follows:
$$
K_{\alpha \beta }\,=\,\left ({{\partial A_{\beta }}\over {\partial \xi ^{\alpha }}}\,-\,
{{\partial A_{\alpha }}\over {\partial \xi ^{\beta }}}\right )\eqno(2.9)
$$
(here the term  connected with a nondifferentiability of the manifold
has not yet been taken into account).
The coefficients $A_{\mu }$ of the form $\omega $ must be obtained either
from the equation of the balance conservation law for energy or from that for
linear momentum. This means that in the first case the coefficients depend
on the energetic action and in the second case they depend on the force action.
In actual processes energetic and force actions have different nature and 
appear to be inconsistent. A commutator of the form $\omega $ constructed of 
derivatives of such coefficients is nonzero.
This means that a differential of the form $\omega $
is nonzero as well. Thus, the form $\omega$ proves to be unclosed.  
This means that the evolutionary relation cannot be an identical one. 
In the left-hand side of this relation it stands a differential  whereas in 
the right-hand side it stands an unclosed form that is not a differential. 

Since the evolutionary relation is not identical, from this relation one 
cannot get the state differential $d\psi$  that may point to the equilibrium 
state of a material system. This means that the material system state is 
nonequilibrium. (The nonequilibrium state means that there is an internal force 
in the material system. It is evident that the internal force originates at the 
expense of some quantity described by the evolutionary form commutator). 
The nonequilibrium state of material system induced by 
the action of internal forces leads to that the accompanying manifolds turns 
out to be a deforming manifold. The metric forms of such manifold cannot be 
closed. The metric form commutator, which describes a deformation of the 
manifold and is nonzero, enters into the commutator of the differential form 
$\omega $ defined on the accompanying manifold. That is, in formula (2.9) it 
will arise the second term connected with the metric form commutator 
with nonzero value. In this case the second term will correlate with the first 
term, and this tern cannot make the differential form $\omega $ 
commutator to be zero. That is, the differential form, which enters into the 
evolutionary equation, cannot become closed. 
And this means that the evolutionary relation cannot become the identical 
relation.  Unclosed differential form $\omega $, which enters into this 
relation, is an example of the evolutionary differential form.

In such a way it can be shown that the evolutionary differential form 
$\omega^p$, involved into this evolutionary relation (2.8), is an unclosed one 
for real processes. Evolutionary relation (2.8) is nonidentical one. 

\subsection*{Obtaining an identical relation from a nonidentical one} 

A role of the nonidentical evolutionary relation in field theory consists in 
that it discloses a connection of the balance conservation law equations 
(equations, which describe physical processes in material media) and the field 
theory equations. Identical relations, which correspond to equations of 
existing field theories, are obtained from the evolutionary nonidentical 
relations, which correspond to the equations describing physical processes 
in material media.
             
The nonidentical relation includes an unclosed differential form. A 
differential of such a form is nonzero. The identical relation includes a closed 
differential form. A differential of such a form equals zero. Hence one can see 
that a transition from the nonidentical relation to the identical one can 
proceed only as a {\it degenerate} transformation.

Let us consider nonidentical evolutionary relation (2.8).

As it has been already mentioned, the evolutionary differential form $\omega^p$,
involved into this relation is an unclosed one for real processes. The
commutator, and hence the differential, of this form is nonzero. That is,
$$
d\omega^p\ne 0\eqno(2.10)
$$
If the transformation is degenerate, from the unclosed evolutionary form it 
can be obtained a differential form closed on pseudostructure. 
The differential of this form equals zero. That is, it is 
realized the transition 

 $d\omega^p\ne 0 \to $ (degenerate transform) $\to d_\pi \omega^p=0$, 
$d_\pi{}^*\omega^p=0$  

{\it The degenerate transformation is realized as a transition from the
accompanying noninertial coordinate system to the locally inertial that}.

To the degenerate transformation it must correspond a vanishing of some 
functional expressions. 
Such functional expressions may be Jacobians, determinants, the Poisson
brackets, residues and others. A vanishing of these functional 
expressions is the closure condition for a dual form. An equality to zero  
of such functional expressions is an identical relation written in terms of 
derivatives (like the Cauchy-Riemann
conditions, canonical relations, the Bianchi identities and so on).
The conditions of degenerate transformation  are connected with symmetries 
that can be obtained from the coefficients of evolutionary and dual forms 
and their derivatives. Since the evolutionary relation has been obtained from 
the equations for material system, it is obvious that the conditions of 
degenerate transformation are specified by properties of the material system. 
The degrees of freedom of material system can correspond to such conditions.
Translational, rotational, oscillatory degrees of freedom are examples.

On the pseudostructure $\pi$ evolutionary relation (2.8) transforms into
the relation
$$
d_\pi\psi=\omega_\pi^p\eqno(2.11)
$$
which proves to be the identical relation. Indeed, since the form
$\omega_\pi^p$ is a closed one, on the pseudostructure it turns
out to be a differential of some differential form. In other words,
this form can be written as $\omega_\pi^p=d_\pi\theta$. Relation (2.11)
is now written as
$$
d_\pi\psi=d_\pi\theta
$$
There are the differentials in the left-hand and right-hand sides of
this relation. This means that the relation is an identical one.
                 								                
Transition from nonidentical relation (2.8) obtained from the 
balance conservation laws to identical relation (2.11) means the following. 
Firstly, it is from such a relation that one can find the state differential 
$d_\pi\psi$. €n existence of the state differential (left-hand side of 
relation (2.11)) points to a transition of  material system to the 
locally-equilibrium state. And, secondly, an emergence of the closed 
(on pseudostructure) inexact exterior form $\omega_\pi^p$  (right-hand side 
of relation (2.11)) points to an origination of the physical structure, namely, 
the conservative object. This object is a conservative physical quantity 
(the closed exterior form  $\omega_\pi^p$) on the pseudostructure (the dual 
form $^*\omega^p$, which defines the pseudostructure).  

From a nonidentical evolutionary relation of degree $p$ (evolutionary relation  
that contains a differential form of degree $p$) one can obtain an identical 
relation of degree $k$, where $k$ ranges from $p$ to $0$.  Under degenerate 
transformation from a nonidentical evolutionary relation one 
obtains a relation being identical on pseudostructure. It is just a relation 
that one can integrate
and obtain a relation with  differential forms of less by one degree.
The relation obtained after integration proves to be nonidentical as well. 
The obtained nonidentical relation of degree $(p-1)$ can be integrated once
again if the corresponding degenerate transformation is realized and the 
identical relation is formed.
By sequential integrating the evolutionary relation of degree $p$ (in the case 
of realization of the corresponding degenerate transformations and forming
the identical relation), one can get closed (on the pseudostructure) exterior 
forms of degree $k$, where $k$ ranges from $p$ to $0$. 

An emergency of identical relation with closed inexact form of degree 
$k$ points to origination of corresponding physical structure.

Thus, a transition from the nonidentical evolutionary relation 
to the identical one elucidates the mechanism of origination of physical structures. 
Such structures form physical fields.

Since the evolutionary relation is obtained from equations for material media, 
it is evident that physical fields are producted by material systems (material 
media). The mechanism of evolutionary processes, which proceed in material 
media and lead to origination of physical structures, has been detailed in 
works [2,3]. In that works it has been 
shown a connection between characteristics of the physical 
structures originated with characteristics of the evolutionary forms, of the 
evolutionary form commutators and of the material system producting these 
structures. In the present work we shall not focus our attention on these 
problems.

Besides, in papers [2,3] it has been shown that parameters, which enter into 
the evolutionary relation, and the identical relations obtained allow to 
classify physical structures and physical fields.

Here it should emphasize the following. 

The evolutionary relation is obtained from not a single, but from several 
equations of the balance conservation laws.
A nonidentity of the evolutionary relation obtained from equations of the
balance conservation laws means that the equations of balance conservation laws 
turn out to be not consistent. And this points to a noncommutativity of the 
balance conservation laws and the nonequilibrium material system state produced 
as a result. A quantity described by the evolutionary differential form
commutator serves as the internal force. A noncommutativity of the balance 
conservation laws is a moving force of evolutionary processes in material 
media. An interaction of the noncommutative balance conservation laws causes 
the evolutionary processes in material media, which lead to origination of 
physical structures.
{\it The noncommutativity of the balance conservation laws
and their controlling role in the evolutionary processes, that are
accompanied by emerging  physical structures, practically
have not been taken into account in the explicit form  anywhere}. The 
mathematical apparatus of evolutionary differential forms enables one to take
into account and describe these points. An account for the noncommutativity 
of the balance conservation laws in material systems enables one to unveil 
the causality of physical processes and phenomena and to understand a meaning 
of postulates that lie at the basis of existing field theories.

\section{Mathematical apparatus of exterior and evolutionary skew-symmetric 
differential forms as the basis of the general field theory}

In section 1 it has been shown that at the basis of the invariant field theories 
there lie the mathematical apparatus of closed exterior differential forms, 
which reflect properties of the conservation laws. 

A connection of field theory with the exterior differential forms 
allow to disclose peculiarities of the field theory equations, their common 
functional properties. From properties of the closed exterior differential forms 
one can see that the field theory equations, the field theory transformations 
and physical interactions are characterized by a degree of the closed form. 
This discloses a relation between them and shows that it is possible to 
introduce a classification of physical fields according to a degree of the 
exterior differential form. Such classification shows that the theory of 
closed exterior differential forms can lie at the basis of the unified field 
theory. 

The field theories that are based on exact conservation laws allow to describe 
the physical fields. However, because these theories are invariant ones 
they cannot answer the question about the mechanism of originating 
physical structures that form physical fields. The origination of physical 
structures and forming physical fields are evolutionary processes,  
and hence they cannot be described by invariant field theories. Only 
evolutionary theory can do this.

As the basis of such evolutionary theory it can be the theory of evolutionary 
skew-symmetric differential forms. It has been shown above that the theory of 
evolutionary skew-symmetric differential forms elucidates a mechanism of 
originating physical structures and forming physical fields and indicates that 
the physical structures, which form physical fields, are producted by material 
systems (material media). The connection of physical fields and material media 
elucidares the causality of physical phenomena and allows to understand what 
specifies the characteristics of physical structures and physical fields. 
Here it should be poined out that (as the present study shows) 
the emergence of physical structures in the evolutionary process proceeds
spontaneously and is manifested as an emergence of certain observable
formations of material system. Such formations and their manifestations are 
fluctuations, turbulent pulsations, waves, vortices, creating massless 
particles, and others.

The evolutionary theory has to be based on properties of the balance 
conservation laws for material systems, because just the interaction of the 
noncommutative balance conservation laws leads to creation of physical 
structures that are generated by material system. 

For developing the evolutionary theory one must know the following.

Firstly, it is necessary to know which material system (medium) generates the 
given physical field. Further, one needs to have an equation that describes 
the balance conservation laws (of energy, linear momentum, angular momentum, 
and mass) for material system [14-16]. After this, it is necessary to get the 
nonidentical evolutionary relation from these equations and to develop 
the method of studying such evolutionary relation by using the 
balance conservation law equations themselves and properties of material 
system (being connected with degrees of freedom). 

The basic mathematical foundations of the theory of evolutionary differential 
forms that describe the evolutionary process in material systems, and the 
mechanism of originating physical structures evidently must be included 
into the evolutionary field theory. However, to realize all 
these mathematical foundations is rather difficult and in many cases this 
turns out to be impossible. 
However, a knowledge of the basic mathematical principles of the theory of 
evolutionary differential forms may be helpful 
while studying a mechanism of originating physical fields. 

The results of qualitative investigations of evolutionary processes on the 
basis of the mathematical apparatus of evolutionary differential 
forms enables one to see the common 
properties that unify all physical fields. The physical fields are generated 
by material media, and at the basis of this it lies the interaction of the 
noncommutative conservation laws of energy, linear momentum, angular momentum, 
and mass for material media. This explains the causality of physical phenomena  
and clarifies the essence of postulates that lie at the basis of existing 
field theories. The postulates, which lie at the basis of the existing field 
theories, correspond to the closure conditions for exterior and dual form, which 
correspond to the conservation laws.

These results allow to classify the physical structures and hence to see 
internal connections between various physical fields. The properties of physical 
structures depend primarily on which material systems (media) generate physical 
structures (but the physical structures generated by different material media 
possess common properties as well). 

In section 2 there were presented parameters according 
to which one can classify physical structures and physical fields. 

One of these 
parameters is, firstly, the evolutionary form degree that enters into the 
evolutionary relation. This is the parameter $p$ that ranges from 0 to 3 
(the case $p=1$ corresponds to interaction of the balance conservation laws of 
energy and linear momentum, the case  $p=2$  does to that of energy, linear 
momentum, and angular momenta, the case  $p=3$  corresponds to interaction of 
the balance conservation laws of energy, linear and angular moment, and mass, 
and to $p=0$ it corresponds an interaction between time and the balance 
conservation law of energy or an interaction between the coordinate and the 
momentum). This parameter specifies a type of physical fields. So, 
the electromagnetic field is obtained from interaction between the balance 
conservation laws of energy and linear and 
angular momenta. The gravitational field is obtained as 
the result of interactions between the balance conservation laws of energy, 
linear momentum, angular momentum, and mass. 

The other parameter is the 
degree of closed differential forms that were realized from given evolutionary 
relation. The values of these parameters designated by $\kappa$ range from 
$p$ to $0$. This parameter, which corresponds to physical structures realized, 
characterizes a connection between physical structures and 
exact conservation laws. A parameter that classifies the equations of invariant 
field theories is such a parameter.

One more parameter is the dimension of space in which the physical 
structures are generated. This parameter points to the fact that the physical 
structures, which belong to common type of the exact conservation laws, 
can be distinguished by their space structure. 

The classification with respect to these parameters may be traced 
in the Table of interactions presented below. In the Table some specific 
features of classification of physical structures were considered.
It will be shown that the classification with 
respect to these parameters not only elucidates connections between  
physical fields generated by material media, but explains a mechanism of 
creating elements of material media themselves and demonstrates  connections 
between material media as well.

In work [2] examples of using the methods developed are presented.

The thermodynamic system has been inspected, and the analysis of the principles 
of thermodynamics has been carried out. It was shown that the first principle 
of thermodynamics is a nonidentical evolutionary relation for thermodynamic
system, and the second principle of thermodynamics is an example of  
identical relation that is obtained from the nonidentical evolutionary relation 
(the first principle of thermodynamics) under realization of the additional 
condition, namely, under realization of the integrating factor that turns 
out to be the inverse temperature. In this case as the closed exterior form it 
serves a differential of entropy. 

Derivation of the evolutionary relation for gas dynamic system is presented. 
The evolutionary relation for gas dynamic system is written for the 
entropy differential as well. But whereas the thermodynamic evolutionary 
relation involves entropy that depends on the thermodynamic parameters, 
the gas dynamic evolutionary relation involves entropy that depends on 
space-time coordinates. It was carried out the investigation of the 
evolutionary form commutator that enters into the gas dynamic evolutionary 
relation. This investigation has shown that the external actions, which 
give contributions into the evolutionary form commutator, effect on 
development of instability and origination of physical structures. This 
analysis allows to understand a mechanism of turbulence. 

It was carried out an analysis of the equations of electromagnetic field. It 
was shown that there are two equations for the Pointing vector from which the 
nonidentical evolutionary relation can be obtained. It was shown under which 
conditions the identical relation follows from that, and this corresponds to 
origination of electromagnetic wave.  

These examples show that the evolutionary approach to field theory enables one 
to get radically new results and to explain the causality of physical phenomena. 

By comparison of the invariant and evolutionary approaches to field theory 
one can state the following. 
Physical fields are described by invariant field theory that is based on exact 
conservation laws. Properties of closed exterior differential forms lie at 
the basis of mathematical apparatus of the invariant theory. A mechanism of 
{\it forming } physical fields can be described only by 
evolutionary theory. The evolutionary theory that is based on 
the balance conservation laws for material systems is just such a theory. 
As the mathematical apparatus of such a theory it can be the mathematical 
apparatus of evolutionary differential forms. It is evident that as the common 
field theory it must serve a theory that involves the basic mathematical 
foundations of the evolutionary and invariant field theories. 

In conclusion we present the Table of data, which can be obtained within the 
framework of the skew-symmetrical differential form theory. The Table shows 
that this theory can be regarded as an approach to the general field theory.
 
\subsection*{\bf Certain classification of physical structures}
As it was shown above, the type of physical structures (and accordingly of 
physical fields) generated by the evolutionary relation depends on the degrees 
of differential forms $p$ and $k$ and on the dimension of initial inertial 
space $n$. Here $p$ is the degree of the evolutionary form 
in the evolutionary relation, which is connected with a number of interacting
balance conservation laws, and $k$ is the  degree of a closed form generated
by the evolutionary relation). By introducing
a classification by numbers $p$, $k$, $n$ one can understand the internal
connection between various physical fields. Since the physical fields are 
carriers of interactions, such classification enables one to see a connection
between interactions. This is reflected in the Table presented below. This 
Table corresponds to elementary particles.

\{It should be emphasized the following. Here the concept of ``interaction"
is used in a twofold meaning: an interaction of the balance conservation laws
that relates to material systems, and the physical concept of ``interaction" 
that relates to physical fields and reflects the interactions of physical 
structures, namely, it is connected with the exact conservation laws\}.

Recall that the interaction of balance conservation laws for energy and
linear momentum corresponds to the value $p=1$, with the balance
conservation law for angular momentum in addition this corresponds to
the value $p=2$, and with the balance conservation law for mass in addition
it corresponds to the value $p=3$. The value $p=0$ corresponds to interaction 
between time and energy or an interaction 
between coordinate and momentum.

In the Table the names of particles created are given. Numbers placed near
particle names correspond to the space dimension. In braces \{\} the
sources of interactions are presented. In the next to the last row we
present the
massive particles (elements of the material system) formed by interactions
(the exact forms of zero degree obtained by
sequential integrating the evolutionary relations with the evolutionary forms
of degree $p$ correspond to these particles). In the bottom row the dimension
of the {\it metric} structure created is presented.

From the Table one can see a correspondence between the degree $k$ of the
closed forms realized and the type of interactions. Thus, $k=0$ corresponds to
the strong interaction, $k=1$ corresponds to the weak interaction,
$k=2$ corresponds to the electromagnetic interaction, and $k=3$ corresponds
to the gravitational interaction.
The degree $k$ of the closed forms realized and the number of interacting
balance conservation laws determine a type of interactions and a type
of particles created. The properties of particles are governed by the space
dimension. The last property is connected with the fact that
closed forms of equal degrees $k$, but obtained from the evolutionary
relations acting in spaces of different dimensions $n$, are distinctive 
because they are defined on pseudostructures of different dimensions
(the dimension of pseudostructure $(n+1-k)$ depends on the dimension 
of initial space $n$). For this reason the realized physical structures
with closed forms of equal degrees $k$ are distinctive in their properties.

\vfill\eject
\centerline{TABLE}

%{\scriptsize

\noindent
\begin{tabular}{@{~}c@{~}c@{~}c@{~}c@{~}c@{~}c@{~}}
\bf interaction&$k\backslash p,n$&\bf 0&\bf 1&\bf 2&\bf 3

\\
\hline
\hline
\bf gravitation&\bf 3&&&&
	\begin{tabular}{c}
	\bf graviton\\
	$\Uparrow$\\
	electron\\
	proton\\
	neutron\\
	photon
	\end{tabular}

\\
\hline
	\begin{tabular}{l}
	\bf electro-\\
	\bf magnetic
	\end{tabular}
&\bf 2&&&
	\begin{tabular}{c}
        \bf photon2\\
	$\Uparrow$\\
	electron\\
	proton\\
	neutrino
	\end{tabular}
&\bf photon3

\\
\hline
\bf weak&\bf 1&&
	\begin{tabular}{c}
	\bf neutrino1\\
	$\Uparrow$\\
	electron\\
	quanta
	\end{tabular}
&\bf neutrino2&\bf neutrino3

\\
\hline
\bf strong&\bf 0&
	\begin{tabular}{c}
	\bf quanta0\\
	$\Uparrow$\\
	quarks?
	\end{tabular}
&
	\begin{tabular}{c}
	\bf quanta1\\
	\\

	\end{tabular}
&
\bf quanta2&\bf quanta3

	\\
\hline
\hline
	\begin{tabular}{c}
	\bf particles\\
	material\\
	nucleons?
	\end{tabular}
&
	\begin{tabular}{c}
	exact\\
	forms
	\end{tabular}
&\bf electron&\bf proton&\bf neutron&\bf deuteron?
\\
\hline
N&&1&2&3&4\\
&&time&time+&time+&time+\\
&&&1 coord.&2 coord.&3 coord.\\
\end{tabular}
%}

(For the value $k=0$  the commutative
relations $\hat q\hat p -\hat p \hat q=\imath \hbar $ correspond to such
quantities. The left-hand side of the commutative 
relations is analog of the commutator value of the nonintegrable
form of zero degree, and the right-hand side is equal to its value at the
instant of realization of the closed zero degree form, the imaginary unit
points to the direction transverse to the pseudostructure).

The parameters $p$, $k$, $n$ can range from 0 to 3. This determine some
completed cycle.
In the Table a single cycle of forming physical structures is presented.
This cycle is related to material system. Each material system has its own
completed cycle. This distinguishes one material system from another system.
One completed cycle can serve as the beginning of another cycle (the structures
formed in the preceding cycle serve as the sources of interactions for
beginning a new cycle). This may mean that one material system (medium)
proves to be imbedded into the other material system (medium). The sequential
cycles reflect properties of sequentially imbedded material systems.

In each cycle one can determine the levels and stages. In the Table presented
rows correspond to the levels and columns correspond to the stages.

From the Table one can see that the cycle level (to which in the Table it 
corresponds the row) points to a type of interaction. This relates to the 
degree $k$ of exterior form realized.

A stage of the cycle (to which in the Table there corresponds a column) is
connected with a total number of the balance conservation laws interacting
in the given space,
namely, with the evolutionary form degree $p$, and with space dimension $n$.
Each cycle involves four stages, to every of which there corresponds
its own value $p$ ($p=0,1,2,3$) and the space dimension $n$.

At each stage of given cycle the transitions from the closed exterior form
of degree $k=p$ to the closed exterior form of degree $k=0$ are the
transitions from one type of interaction to another. Such transitions
execute the connection between different types of interactions.

At each stage the transition from the closed inexact form of zero degree
$k=0$ to the exact exterior form of the same degree corresponds to the
transition from relevant physical structure to the element of material system.
To each type there corresponds its own appropriate coupling constant.
This means that from the physical structure it can be obtained the appropriate
elements of material system.  In every cycle four types
of elements that are distinguished by the dimensions  of their metric
structure are created. In the Table presented  electron, proton, neutron,
and deuteron(?) are such elements.

Each stage has the specific features that are inherent to the same stages
in other cycles. $\{$One can see this, for example, by comparison of the cycle
described with the other cycle, where to the exact form there correspond 
sequentially conductors, semiconductors, dielectrics, and neutral elements. The
properties of elements of the third stage, namely, neutrons
in one cycle and dielectrics in the other coincide with those of the so-called
"magnetic monopole" [17,18]$\}$.
Physical structures that have the same parameters exhibit common properties.
And yet the physical structures that have the same parameters
$p,\,k,\,n$ will be distinctive according to in what cycle they are located.
That is, which material system generates these structures. (As it was already
pointed out, thermodynamic, gas dynamic, cosmic systems, the system of 
elementary particles and so on can serve as examples of material system. 
The physical vacuum in its properties may be regarded as an analog of a material 
system that generates some physical fields.).

The Table presented provides the idea about the dimension of pseudostructures
and  metric structures.

It was shown that the evolutionary relation
of the degree $p$ can generate (with the availability of degenerate
transformations)  closed forms of degrees $0\le k\le p$ on the 
pseudostructures.
Under generation of the forms of sequential degrees $k=p$, $k=p-1$, \dots, 
$k=0$ the pseudostructures of the dimensions $(n+1-k)$: 1, \dots, $n+1$,
where $n$ is the dimension of initial inertial space, are obtained.
While going to the exact exterior form
of zero degree the metric structure of the dimension $N=n+1$ is obtained. With 
a knowledge of the values $n$ and $k$ for each physical structure presented 
in the Table one can find the dimension of relevant pseudostructure.
In the bottom row of the Table the dimension $N$ of the metric structure 
formed is presented.  From initial space of the dimension $0$ the metric space 
of the dimension $1$ (it can occurs to be time) can be realized. From space of
the dimension $1$ the metric space of dimension $2$ (time and 
coordinate) can appear and so on. From initial space of the dimension $3$
it can be formed the metric space of dimension $4$ (time and $3$
coordinates). Such space is convolved  and a new dimension cannot already be
realized. This corresponds to ending the cycle. (Such metric space with
corresponding physical quantity that is defined by the exact exterior form
is the element of new material system.)

1. Petrova L.~I., Origination of physical structures. //Izvestia RAN, Fizika, N 1, 2003.  

2. Petrova L.~I., Exterior and evolutionary skew-symmetric differential forms 
and their role in mathematical physics. http://arXiv.org/pdf/math-ph/0310050

3. Petrova L.~I., Conservation laws. Their role in evolutionary processes.   
(The method of skew-symmetric differential forms). 
http://arXiv.org/pdf/math-ph/0311008

4. Bott R., Tu L.~W., Differential Forms in Algebraic Topology. 
Springer, NY, 1982.

5. Encyclopedia of Mathematics. -Moscow, Sov.~Encyc., 1979 (in Russian).

6. Novikov S.~P., Fomenko A.~P., Elements of the differential geometry and 
topology. -Moscow, Nauka, 1987 (in Russian).  

7. Cartan E., Lecons sur les Invariants Integraux. -Paris, Hermann, 1922.   

8. Schutz B.~F., Geometrical Methods of Mathematical Physics. Cambrige 
University Press, Cambrige, 1982.

9. Wheeler J.~A., Neutrino, Gravitation and Geometry. Bologna, 1960.

10. Smirnov V.~I., A course of higher mathematics. -Moscow, Tech.~Theor.~Lit. 
1957, V.~4 (in Russian).

11. Tonnelat M.-A., Les principles de la theorie electromagnetique 
et la relativite. Masson, Paris, 1959.

12. Pauli W. Theory of Relativity. Pergamon Press, 1958.

13. Weinberg S., Gravitation and Cosmology. Principles and applications of 
the general theory of relativity. Wiley \& Sons, Inc., N-Y, 1972.

14. Clark J.~F., Machesney ~M., The Dynamics of Real Gases. Butterworths, 
London, 1964.

15. Fock V.~A., Theory of space, time, and gravitation. -Moscow, 
Tech.~Theor.~Lit., 1955 (in Russian).

16. Tolman R.~C., Relativity, Thermodynamics, and Cosmology. Clarendon Press, 
Oxford,  UK, 1969.

17. Dirac P.~A.~M., Proc.~Roy.~Soc., {\bf A133}, 60 (1931).

18. Dirac P.~A.~M., Phys.~Rev., {\bf 74}, 817 (1948). 

\end{document}